\documentstyle[12pt,epsfig]{article}  
\begin{titlepage}
\begin{document}
\noindent
DESY 00-188  \hspace*{8.5cm} ISSN 0418-9833 \\
December 2000

\vspace*{3cm}

\begin{center}
  \Large
{\bf $xF_3^{\gamma Z}$ in Charged Lepton Scattering}

\vspace*{1.2cm}
    {E.Rizvi$^{(a)}$ and T.Sloan$^{(b)}$}   \\
  \                                         \\
  \  $(a)$ School of Physics and Space Research, University of Birmingham, 
UK.\\
  \  $(b)$ Dept of Physics, University of Lancaster, UK. 
\end{center}
\vspace*{0.5cm}

\begin{abstract}
\noindent
Measurements of the structure function $xF_3^{\gamma Z}$ are now becoming 
available in charged lepton deep inelastic scattering at HERA.  The 
correction factors which are necessary to compare these with measurements 
of $xF_3^{\nu N}$ in neutrino-nucleon scattering are deduced and a comparison 
is made between the H1 and the CCFR data.  A sum rule is derived 
for the structure function $xF_3^{\gamma Z}$ measured in charged lepton 
scattering which is the analogue of the Gross Llewellyn-Smith sum rule 
for neutrino-nucleon scattering. 
\end{abstract}
\vfill

\begin{center}
 {Scientific Note submitted to the European Physical Journal C}
\end{center}
\end{titlepage}

\pagestyle{plain}

\section{Introduction}

At large squared four-momentum transfer, $Q^2$, the effects 
of $Z^0$ and $W^\pm$ exchange become significant in charged lepton deep 
inelastic scattering relative to the single photon exchange process 
which dominates at lower values of $Q^2$. At such large values of 
$Q^2$, following the notation in \cite{H1}, the cross section for 
unpolarised neutral current deep inelastic scattering, 
$e^\pm p \rightarrow e^\pm X$, after correction for QED radiative 
effects is related to the structure functions of the proton by 
\begin{equation}
\frac{d^2\sigma^\pm_{NC}}{dx dQ^2} = \frac{2 \pi \alpha^2}{x Q^4}
(1+\delta^{weak}_{NC})[Y_+ \tilde{F}_2 \mp Y_-x\tilde{F}_3 - y^2 \tilde{F}_L]. 
\label{main}
\end{equation}
Here $\alpha$ is the fine structure constant, $\delta^{weak}_{NC}$
are the weak radiative corrections described in \cite{Hub1}, the helicity 
dependences of the electroweak interactions are contained in the 
functions $Y_\pm = 1 \pm (1-y)^2$ and $x$ and $y$ are the 
Bjorken variables. The generalised structure functions 
$\tilde{F}_2$ and $x\tilde{F}_3$ can be decomposed as follows
\begin{equation}
\tilde{F}_2 = F_2 - v_e \frac{\kappa_w Q^2}{Q^2+M_Z^2} F_2^{\gamma Z} 
+(v_e^2+a_e^2) (\frac{\kappa_w Q^2}{Q^2+M_Z^2})^2 F_2^Z
\end{equation}
\begin{equation} 
x\tilde{F}_3 =~~ - a_e \frac{\kappa_w Q^2}{Q^2+M_Z^2} xF_3^{\gamma Z} 
+(2v_ea_e) (\frac{\kappa_w Q^2}{Q^2+M_Z^2})^2 xF_3^Z,
\end{equation}  
where $\kappa_w^{-1} = 4 M_W^2/M_Z^2~(1 - M_W^2/M_Z^2)$ in the on-shell 
scheme \cite{PDG} where $M_W$ is defined in terms of the electroweak 
inputs. Here $M_Z$ and $M_W$ are the masses of the $W^\pm$ and $Z^0$ 
vector bosons.     
The quantities $v_e$ and $a_e$ are the vector and 
axial vector couplings of the electron to the $Z^0$ and are related to 
the weak isospin of the electron, namely $v_e=-1/2+2 \sin^2\theta_w$ and 
$a_e=-1/2$ \cite{PDG} where $\theta_w$ is the electroweak mixing angle. 
The electromagnetic structure function $F_2$ originates 
from photon exchange only. The functions $F_2^Z$ and $xF^Z_3$ are the 
contributions to $\tilde{F}_2$ and $x \tilde{F}_3$ from $Z^0$ exchange 
and the functions $F_2^{\gamma Z}$ and $F_3^{\gamma Z}$ are the 
contributions from the interference of the $\gamma$ and $Z^0$ exchange 
amplitudes.  The longitudinal structure 
function $\tilde{F}_L$ may be similarly decomposed. 

The structure function $x \tilde{F}_3$ can be measured in charged 
lepton deep inelastic scattering by taking the difference between 
the measured cross sections for electron and positron scattering, 
as described in \cite{H1}. 
In this way the major contributions from $\tilde{F}_2$ 
and $\tilde{F}_L$ are eliminated. In the 
kinematic range of HERA the contribution of 
$xF_3^Z$ is small and can be safely neglected.  

In the quark parton model of the nucleon the structure functions 
$xF_3^{\gamma Z}$ and $xF_3^Z$ are related to the electroweak 
couplings and an incoherent sum of the parton distribution functions 
(PDFs) 
\begin{equation}
[xF_3^{\gamma Z},xF_3^Z]=x\sum_q [2e_q a_q,2v_qa_q](q-\bar{q}) 
\label{epxf3}
\end{equation}
where $q$,$\bar{q}$ are the quark PDFs, $e_q$ is the charge of the quark,   
$v_q=I_{3q}-2e_q \sin ^2 \theta_w$ and $a_q=I_{3q}$ are the quarks' 
electroweak couplings with $I_{3q}$ the weak isospin of the quark. 
In neutrino nucleon scattering, averaging 
over the scattering from neutron and proton targets and assuming from 
isospin conservation that the valence $u$ quarks in the neutron 
have the same distribution as the valence $d$ quarks in the proton 
and vice versa, $xF_3^{\nu N}$ has a similar relation, of the form 
\begin{equation}
xF_3^{\nu N}(x,Q^2) = \sum_q x(q(x,Q^2)-\bar{q}(x,Q^2)). 
\label{nupxf3}
\end{equation}

\section{Comparison of Charged Lepton and Neutrino Data.}

It can be seen from equations (\ref{epxf3}) and (\ref{nupxf3}) 
that the measured value of $xF_3^{\nu N}$ in $\nu$-Nucleon and 
$xF_3^{\gamma Z}$ in charged lepton scattering are  
comparable within the Quark-Parton model after allowance is made 
for the different targets and electroweak couplings.  
Both are related to the differences of 
the PDFs for quarks and antiquarks which are dominated by 
the valence quarks of the target nucleon. 
Taking the ratio of equations  (\ref{epxf3}) and (\ref{nupxf3}), 
substituting the $u$ and $d$ quark charges and couplings, gives 
\begin{equation}
\frac{xF_3^{\gamma Z}(x,Q^2)}{xF_3^{\nu N}(x,Q^2)} 
= \frac{2}{3} \left( \frac {1+\frac{1}{2}\frac{d_v}{u_v}}{1+\frac{d_v}{u_v}}
\right)
\label{rati2}
\end{equation}
where the quark-antiquark differences have been replaced by the valence
distributions, $u_v$ and $d_v$. 

Hence it is possible to compare the $\nu$ and charged lepton data by 
correcting one of them by the factor on the right hand side of equation 
(\ref{rati2}). The correction factor for the quark distributions is 
calculable and is not very sensitive to the difference between the 
$u_v$ and $d_v$ distributions. For example, if one makes the naive 
assumption that $d_v/u_v$ is 1/2 then 
\begin{equation}
\frac {1+\frac{1}{2}\frac{d_v}{u_v}}{1+\frac{d_v}{u_v}} \sim 0.83. 
\end{equation}
Taking the values of $d_v$ and $u_v$ from the latest parton
distribution functions \cite{MRS99} this correction factor varies from
0.811 at the lowest $x$ to 0.921 at the highest $x$ values. If 
instead we use the fit to the H1 data \cite{H1F2} the factor varies 
from 0.826 to 0.923 over the same range of $x$. Furthermore, the 
$Q^2$ dependence of this factor is very weak since there is a
cancellation from the ratio $d_v/u_v$. 
Hence there is 1-2$\%$  
uncertainty coming from the quark distribution functions in evaluating 
the correction factor to compare neutrino data on $xF_3^{\nu N}$ and 
charged lepton data on $xF_3^{\gamma Z}$. 

\begin{figure} 
\vspace{2mm}
\hspace*{-7mm}\epsfig{file=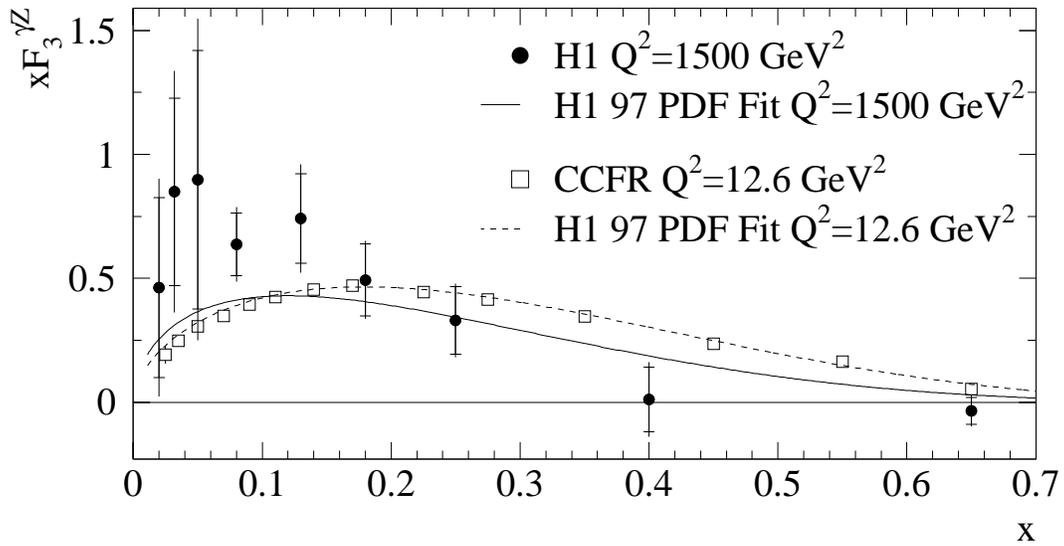,width=15cm}
\caption[junk]{ The H1 data on $xF_3^{\gamma Z}$ compared with the CCFR 
data at $Q^2$=12.6 GeV$^2$ multiplied by the ratio given in equation
(\ref{rati2}). This ratio was computed from the values of $u_v$ and $d_v$ 
taken from the QCD fit described in \cite{H1F2}. The smooth curves 
(labelled H1 97 PDF Fit) are also calculated from this fit.  } 
\label{rati1}
\end{figure}

Figure (\ref{rati1}) shows the H1 measurement \cite{H1} of 
$xF_3^{\gamma Z}$ after introducing the small correction for the 
scaling violations determined from the H1 97 PDF Fit~\cite{H1F2}. 
Weighted means have been taken of the measurements corrected 
to $Q^2$ = 1500 GeV$^2$  at the same $x$ value. The smooth curves show the
expected variation from the fit to fixed target and H1 data. The H1 data
are compared to the CCFR measurements of $xF_3^{\nu N}$ \cite{CCFR} after
correction by the ratio computed from eq. \ref{rati2}.   

The difference between the curves at $Q^2$=12.6 GeV$^2$ and 
$Q^2$=1500 GeV$^2$ indicates 
the softening of the structure function expected from the DGLAP
evolution for a non-singlet structure function.  

\section{A Sum Rule for $xF_3^{\gamma Z}$ in Charged Lepton Scattering}

The structure function $F_3^{\gamma Z}$ in charged lepton scattering 
should obey the sum rule  
\begin{equation}
\int_0^1 F_3^{\gamma Z} dx = 
(2 e_u a_u N_u + 2 e_d a_d N_d)~{\cal O}(1-\frac{\alpha_s}{\pi}) 
      = \frac{5}{3}~{\cal O}(1-\frac{\alpha_s}{\pi}).   
\label{sumrule}
\end{equation}
This follows from replacing the difference between the quark
distributions in equation (\ref{epxf3}) by the valence quark
distributions. The fact that there are $N_u$=2 $u$ and $N_d$=1 
$d$ valence quarks in the proton allows the integrals over these 
distributions to be replaced by $N_u$ and $N_d$. This gives the 
prediction in equation (\ref{sumrule}) to which QCD radiative 
corrections ${\cal O}(1-\alpha_s/\pi) \sim -5\%$ should be applied \cite{Rus}. 
A similar sum rule (the Gross Llewellyn-Smith sum rule \cite{GLS}) has 
been observed to be valid for neutrino data\cite{CCFR2}. Evaluating the 
integral over the H1 data gives
\begin{equation}
\int_{.02}^{.65} F_3^{\gamma Z} dx = 1.88 \pm 0.35(stat.) \pm 0.27(sys.).
\end{equation}
The value expected from the standard model (obtained by integrating 
over the solid curve in Fig \ref{rati1})
is $\int_{.02}^{.65} F_3^{\gamma Z} dx = 1.11$ and over the 
whole range is  $\int_{0}^{1} F_3^{\gamma Z} dx = 1.59$. Here the difference 
of the latter from 5/3 arises from the QCD radiative corrections.   
The H1 measurement is 1.7 standard deviations above the value expected 
from the standard model. 

\section{Discussion of the Result}

Fig (\ref{rati1}) indicates that the H1 data on $xF_3^{\gamma Z}$ have a 
tendency to become softer than expected from pure DGLAP evolution 
for a non-singlet structure function. The sum rule integral described in 
the previous section shows that the effect is significant at roughly 
the level of 1.7 standard deviations and it could be 
within the experimental error. The future high luminosity running at 
HERA is needed to study the $x$ dependence of $xF_3^{\gamma Z}$ 
in greater detail.   

The radiative corrections were 
examined closely to see if they could contribute a systematic effect. 
A difference between $e^+$ and $e^-$ scattering
\cite{refute} is expected from the interference of two photon exchange 
with the single photon exchange diagram. Such corrections are 
implemented in the HERACLES generator but not in the DJANGO Monte Carlo 
used to correct the data \cite{Hub2}. Studies of the HERACLES 
generator showed that this effect tends to increase all the values 
of $xF_3^{\gamma Z}$ derived from the data by a magnitude which 
is small compared to the size of the errors. Hence it is valid 
to neglect such corrections in the present data. It may become 
necessary to include these corrections as the data become more precise 
following the luminosity upgrade of HERA.

\section{Conclusion}

The new measurements by H1 of $xF_3^{\gamma Z}$ in charged lepton 
neutral current deep inelastic scattering at HERA have been discussed 
and compared to the neutrino data on $xF_3^{\nu N}$. 
A new sum rule is described which governs the integral 
of $F_3^{\gamma Z}$ and the H1 data are compatible with this sum rule 
within 1.7 standard deviations. Hence, although the H1 data tend to 
be systematically above the prediction of the standard model, the 
difference from the model is within the experimental precision.   
High precision measurements of $xF_3^{\gamma Z}$ will become possible 
with the increased luminosity soon to be available at HERA,   
allowing this structure function to be studied in greater detail.    

\section{Acknowledgement}

This work was carried out within the framework of the H1 Collaboration and  
we thank all our colleagues in the collaboration for their help and support. 
We especially wish to thank Eckhard Elsen, Dieter Haidt and Yves Sirois 
for their knowledgeable comments. We are very grateful to Hubert 
Spiesberger for invaluable and helpful discussions.

\end{document}